\documentclass[twocolumn,prb,amsmath,amssymb,amsfonts,superscriptaddress,floatfix,showpacs]{revtex4}

\usepackage{graphicx}
\usepackage{dcolumn}
\usepackage{bm}
\usepackage{rotating} 

\begin{document}

\title{Implementation of an all-electron GW approximation based
on the PAW method without  plasmon pole approximation: application to Si, SiC, AlAs, InAs, NaH
and KH}
\date{\today}
\author{S.~Leb\`egue}
\affiliation{Institut de Physique et de Chimie des Mat\'eriaux de
Strasbourg (IPCMS), UMR 7504 du CNRS, 23 rue du Loess, 67037 Strasbourg, France, EU}
\affiliation{Department of Physics, University of California, Davis, California 95616 }
\author{B.~Arnaud}
\affiliation{Groupe Mati\`ere condens\'ee et Mat\'eriaux
(GMCM), Campus de Beaulieu - Bat 11A
35042 Rennes cedex, France, EU}
\author{M.~Alouani}
\affiliation{Institut de Physique et de Chimie des Mat\'eriaux de
Strasbourg (IPCMS), UMR 7504 du CNRS, 23 rue du Loess, 67037 Strasbourg, France, EU}
\affiliation{Kavli Institute of Theoretical Physics, University of California, Santa Barbara, Ca 93111}
\author{P.~E.~ Bloechl}
\affiliation{Kavli Institute of Theoretical Physics, University of California, Santa Barbara, Ca 93111}
\affiliation{Institute of Theoretical Physics, Clausthal University of Technology, Leibnizstr. 10
D-38678 Clausthal Zellerfeld,  Germany, EU}

\begin{abstract}
A new  implementation of the GW approximation (GWA) based on the
all-electron  Projector-Augmented-Wave method (PAW) is presented,  
where the screened Coulomb interaction is computed within the Random Phase 
 Approximation (RPA) instead of  the  plasmon-pole model.
Two different ways of computing the self-energy are reported. The method is used
successfully to determine the quasiparticle energies of 
six semiconducting or insulating materials: Si,
 SiC, AlAs, InAs, NaH and KH. 
To illustrate the novelty of the method the real and imaginary part of the 
frequency-dependent self-energy together with  
the spectral function of silicon are computed.  Finally, 
the GWA results are compared  with other 
calculations, highlighting  that all-electron  GWA  results  can differ 
markedly from those based on pseudopotential approaches.
\end{abstract}


\maketitle
\section{\label{sec:one} Introduction}
For many weakly correlated materials, the density-functional theory\cite{Hohenberg}(DFT) in 
the local-density approximation (LDA) provides a good description 
of their ground-state properties. However, DFT is not able to
 describe correctly their excited states. 
Thus, for example, the band gaps in the
 LDA are  typically much smaller than  the experimental values. Quasiparticle (QP)
 electronic-structure  calculations beyond the DFT  are  therefore highly desirable. \\
\indent The GW approximation (GWA) of Hedin\cite{Hedin1,Hedin2}, which produces a good approximation
 for the electron's self-energy $\Sigma$, enables us  to make 
 first-principle QP calculations for realistic  materials. Thus, the GWA
 has been successfully applied to the calculation of QP
 electronic  structures of many kinds of materials\cite{review_gw1,review_gw2,review_gw3,review_gw4}.
 In particular, recent success has been achieved  on predicting the metal-insulator transition
 of bcc hydrogen,\cite{blase-hydro} electronic excitations of yttrium trihydride,\cite{gelderen-ytt}
as well as the QP electronic structure of copper.\cite{marini-copper}
 Unfortunately, most of the GWA implementations are based on the 
pseudopotential type of approaches 
 together with  plasmon-pole (PlP) models.\cite{Hybert_86,Godby_old,Rohlfing,Hamada,brice,furth-big}
 The weakness of these types of  calculations  is
 that the imaginary part of the self-energy is not accessible,
  making it impossible to determine spectral functions and hence 
to interpret photoemission experiments. In addition 
the PlP approximation is expected not to hold for systems with localized
electrons.  Moreover, it has been noticed recently\cite{brice,kotani}
 that GWA implementations based on  pseudopotential methods lead 
 to larger and more $\bf{k}$-dependent shifts than calculations based 
 on all-electron DFT methods, bringing into question  the validity of the former approaches.\\
\indent However, some attempts have been  made to go beyond the plasmon-pole 
approximation.\cite{review_gw4,marini-copper,kotani,wei,Godby,bechstedt,fleszar} 
In particular, Aryasetiawan has approximately determined the screening within the RPA using a linear muffin-tin
orbital method within the atomic sphere approximation\cite{review_gw4}  ( LMTO-ASA).
This method, although fast, approximates the space by atomic centered overlapping spheres, 
thus completely neglecting the interstitial region, and hence making the reliability of the GW method uncertain. 
Kotani and van Schilfgaarde based their full-potential LMTO GW calculation\cite{kotani} 
on the work of Aryasetiawan by taking into account correctly the interstitial region. 
Nevertheless their method is not quite accurate since in their implementation they 
didn't take into account the multiplicity of the same angular momenta for a given  principal 
quantum number in the basis set (like simultaneously using the 3d  and 4d states).
Finally, Ku and Eguiluz produced  selfconsistent and non-selfconsistent 
QP band gaps based on  an approximate Luttinger-Ward 
functional,\cite{wei} the non-selfconsistent results  are much smaller than 
all existing GW calculations. Since these results are based on 
a different scheme we have chosen  not  to discuss their method  further.
On the other hand, several pseudopotential 
have produced GW results without resorting to the
plasmon-pole approximation.  These methods, although interesting, 
use pseudowave 
functions and hence  can only determine  pseudo-matrix elements of operators,  
 making them difficult to justify as  quantitative and reliable methods for 
computing 
QP  properties. \\
\indent The major purpose of this paper is then to present a new  implementation
 of the GWA method using the all-electron full-potential Projector Augmented-Wave method (PAW)
{\sl complete} basis set, and  without  using any PlP model for the  determination of the
 dielectric function. The screening of the Coulomb
 interaction is thus described in the Random Phase Approximation (RPA), 
 avoiding further approximations.\\
\indent The paper is organized as follows. In section \ref{sec:two} we
describe our implementation of the GW approximation. In Section \ref{sec:three}
 we present our QP calculations for Si, SiC, AlAs, and InAs and also for the
 alkali hydrides compounds NaH and KH (to our knowledge this is the first
GWA study of these alkali hydrides).  At the end of this section we  compare
 and discuss our results with other calculations and experiments. 
\section{\label{sec:two} Formalism}

\subsection{The PAW method} 
The PAW formalism has been well-described 
elsewhere,\cite{Bloechl,vasp,holzwarth,holzwarth2} so we 
will not discuss it in this paper. 
The PAW method\cite{Bloechl} is a very powerful all-electron method for performing electronic structure calculations
 within the framework of the LDA. It takes advantage of   
the simplicity of pseudopotential methods, but describes correctly the  nodal behavior 
in the augmentation regions.  The selfconsistent calculation of the electronic 
structure is performed using the Car-Parinello method over the occupied states.
To determine the eigenvalues and eigenvectors of all unoccupied states (up to 200 eV above 
the top of the valence states)  needed for the GW calculations, we have extracted the selfconsistent
full-potential, constructed and diagonalized the PAW Hamiltonian for every irreducible {\bf k} point in 
the Brillouin zone.

\subsection{The GW approximation} 

\subsubsection{Quasiparticle energies} 
In general, the QP energies $E_n({\bf k})$ and wave function $\psi_{{\bf k}n}({\bf r})$
are determined from the solution of the QP equation \\

$(T+V_{ext}+V_{h})\psi_{{\bf k}n}({\bf r}) + \int d^3r^{\prime}
 \Sigma({\bf r},{\bf r}^{\prime},E_n({\bf k}))\psi_{{\bf k}n}({\bf r}^{\prime}) $
\begin{eqnarray}
= E_n({\bf k})\psi_{{\bf k}n}({\bf r}) ~~~~~~~~
\label{eq:qp_psi}
\end{eqnarray}
where $T$ is  the free-electron kinetic energy operator, $V_{ext}$ the external potential 
 due to the ion cores, $V_{h}$ the average electrostatic (Hartree) potential,
 and $\Sigma$ the electron self-energy operator. 
The major difficulty connected with Eq. (\ref{eq:qp_psi}) is finding 
  an adequate approximation for the self-energy operator
 $\Sigma({\bf r},{\bf r}^{\prime},E_n({\bf k}))$.
Nonetheless, it was shown by Hedin\cite{Hedin1,Hedin2} that  writing 
the self-energy 
as a product of the Green's function and the screened Coulomb interaction $W$
yields the successful  GW approximation for $\Sigma$.
In this approximation,  both the non-locality and the dynamical correlations are included.
Assuming that the difference $\hat{\Sigma}-\hat{V}_{xc}$ between the self-energy and the Kohn-Sham
exchange and correlation potential is small, we can use a perturbation theory approach to solve the 
effective QP Hamiltonian $\hat{H}^{qp}$
\begin{equation}
\hat{H}^{qp} = \hat{H}_{KS} + (\hat{\Sigma}-\hat{V}_{xc})
\label{eq:hqp}
\end{equation}
and determine the QP energies  by expanding the real part of selfenergy  to first order  around 
$\epsilon_n({\bf k})$ thus making the comprison with the PlP models possible \\
$\textrm{Re}E_n({\bf k})) = \epsilon_n({\bf k})+ Z_{n{\bf k}} \times$
\begin{eqnarray}\label{quasiparticle_energy_final}
[\langle\Psi_{{\bf k}n}|
\textrm{Re}\Sigma({\bf r},{\bf r}^{\prime},\epsilon_n({\bf k}))|\Psi_{{\bf k}n}\rangle
- \langle\Psi_{{\bf k}n}|V_{xc}^{LDA}(r)|\Psi_{{\bf k}n}\rangle]
\end{eqnarray}
where the QP renormalization factor $Z_{n{\bf k}}$ is given by
\begin{equation}\label{Renormalization}
Z_{n{\bf k}}=[1-\langle\Psi_{{\bf k}n}|
\frac{\partial}{\partial\omega} \textrm{Re}\Sigma({\bf r},{\bf r}^{\prime},
\epsilon_n({\bf k}))
|\Psi_{{\bf k}n}\rangle]^{-1}.
\end{equation}
This assumption is valid for simple $sp$ bonded materials, since it was 
shown that  the QP wave function $\psi_{{\bf k}n}$ and Kohn-Sham
 wave function $\Psi_{{\bf k}n}$ are almost 
identical, i.e., the QP Hamiltonian  $\hat{H}^{qp}$ is diagonal in 
the $\Psi_{{\bf k}n}$ basis for
simple $sp$ bonded semiconductors.\cite{Godby_old,Hybert_86} We therefore assume 
this behavior for the materials studied in this paper. According to this
equation, the LDA eigenvalues $\epsilon_n({\bf k})$ are then corrected by
the GW approximation.  
The numerical work is  therefore considerably reduced, but still 
computationally demanding.  \\
\indent 
In our implementation, we have calculated the Green's
function only for the valence and conduction states. One has then 
to subtract out only the valence exchange and correlation potential in Eq. 
(\ref{quasiparticle_energy_final}).
 To check the accuracy of this procedure,
we have also used the so-called Hartree-Fock decoupling,\cite{massidda,brice} and have
found that the average error in  the QP energies  of Si 
with respect to the top of the valence states is 0.05 eV.
The approximation made here is the one currently used in all 
pseudopotential-based GWA calculations, making our method compatible with existing GW implementations.

\subsubsection{Screened Coulomb interaction} 
For the calculation of the self-energy, one needs to evaluate the
 dynamically screened interaction $W({\bf r},{\bf r}^{\prime},\omega)$,
 which can be  rewritten in reciprocal space as:
\begin{equation}\label{fourier_screened_interaction}
W_{{\bf G},{\bf G}^{\prime}}({\bf q},\omega)=4\pi\frac{1}{|{\bf q}+{\bf G}|}
\tilde{\epsilon}^{-1}_{{\bf G},{\bf G}^{\prime}}({\bf q},\omega)
\frac{1}{|{\bf q}+{\bf G}^{\prime}|}.
\end{equation}
The symmetrized dielectric matrix 
$\tilde{\epsilon}_{{\bf G}{\bf G}^{\prime}}({\bf q},\omega)$ is defined 
in the random phase approximation (RPA) by\cite{adler}
\begin{widetext}
\begin{eqnarray}
\label{symetrised_epsilon}
\tilde{\epsilon}_{{\bf G}{\bf G}^{\prime}}({\bf q},\omega) & = & \delta_{{\bf G}{\bf G}^{\prime}}
-\frac{8\pi}{\Omega|{\bf q}+{\bf G}||{\bf q}+{\bf G}^{\prime}|}
\sum_{v,c,{\bf k}} M^{vc}_{{\bf G}}({\bf k},{\bf q}) 
[M^{vc}_{{\bf G}^{\prime}}({\bf k},{\bf q})]^{*} \nonumber \\
& & \times \left(\frac{1}{\omega +\epsilon_{v}({\bf k}-{\bf q})-\epsilon_{c}({\bf k})-i\delta} -
\frac{1}{\omega-\epsilon_{v}({\bf k}-{\bf q})+\epsilon_{c}({\bf k})+i\delta}\right), 
\end{eqnarray}
\end{widetext}
with the following notation:
\begin{equation} \label{matrix_element}
M^{nm}_{{\bf G}}({\bf k},{\bf q})=
\langle\Psi_{{\bf k}-{\bf q}n}|e^{-i({\bf q}+{\bf G}){\bf r}}|\Psi_{{\bf k}m}\rangle, 
\end{equation}
where $v$ and $c$  denote, respectively, the  valence and conduction states, and 
$\delta$ a positive infinitesimal. 
The matrix elements given by Eq. (\ref{matrix_element}) are evaluated
 using the PAW basis set as described in Ref. \onlinecite{brice}.

Most of GWA calculations use a kind of PlP approximation. This is 
  computationally efficient since one  obtains an analytic expression 
 for the integral in the self-energy.
It is not clear, however,  that this kind of approximation is valid 
for describing the QP of different kind of materials. It is for this 
reason that we have chosen to avoid the PlP model altogether,  to  compute 
the dynamical dielectric function in the RPA 
(\ref{symetrised_epsilon}),  and to perform the integral of the selfenergy
numerically.  In our implementation, we need to compute 
$\tilde{\epsilon}_{{\bf G}{\bf G}^{\prime}}({\bf q},\omega)$ along the 
imaginary axis and for some real frequencies. This technical point will become
clearer in the next subsection.  \\
\indent To reduce the computational cost of the
GWA, we use symmetry properties. Details about the utilization
 of the symmetry for the static dielectric matrix has been already given elsewhere\cite
{Hybert_87,review_gw2,aulbur_thesis,brice},
 so we just describe briefly  how to use the symmetry in the case of 
the dynamical dielectric function.
For the case of pure imaginary frequencies, we could safely ignore 
 the broadening factor $i\delta$; in this case 
$\tilde{\epsilon}_{{\bf G}{\bf G}^{\prime}}({\bf q},i\omega)$ is hermitian
 and we could use the symmetry just as  in the static case.
We can then write the symmetrized dielectric matrix as 
\begin{widetext}
\begin{eqnarray} \label{imaginary_sym}
\tilde{\epsilon}_{{\bf G}{\bf G}^{\prime}}({\bf q},i\omega) & = & \delta_{{\bf G}{\bf G}^{\prime}}
-\frac{8\pi}{\Omega|{\bf q}+{\bf G}||{\bf q}+{\bf G}^{\prime}|}
\sum_{{\bf k}\in BZ_{{\bf q}}}\sum_{v,c} \sum_{R \in G_{\bf q}}
M^{vc}_{R{\bf G}}({\bf k},{\bf q})[M^{vc}_{R{\bf G}^{\prime}}({\bf k},{\bf q})]^{*} \nonumber \\
& & \times \left(\frac{1}{i\omega + \epsilon_{v}({\bf k}-{\bf q})-\epsilon_{c}({\bf k})} -
\frac{1}{i\omega - \epsilon_{v}({\bf k}-{\bf q})+\epsilon_{c}({\bf k})}\right),
\end{eqnarray}
\end{widetext}
where $G_{\bf q}$ is the little group of the point group $G$ such that $R{\bf q}={\bf q}$;
$R$ being a symmetry operation.
The computational cost is further  reduced by noticing that
\begin{equation} \label{imaginary_sym2}
\tilde{\epsilon}_{{\bf G}{\bf G}^{\prime}}(R{\bf q},i\omega)
=\tilde{\epsilon}_{R^{-1}{\bf G}R^{-1}{\bf G}^{\prime}}({\bf q},i\omega).
\end{equation}
for real $\omega$, although the dielectric matrix is not hermitian,
 we could use the symmetry by making a decomposition into  hermitian and
 anti-hermitian parts of the polarizability $P_{{\bf G}{\bf G}^{\prime}}({\bf q},\omega)$. 
if we define $A_{{\bf G}{\bf G}^{\prime}}({\bf q},\omega)$
 and $B_{{\bf G}{\bf G}^{\prime}}({\bf q},\omega)$ by 
\begin{eqnarray}
A_{{\bf G}{\bf G}^{\prime}}({\bf q},\omega) = 
\frac{P_{{\bf G}{\bf G}^{\prime}}({\bf q},\omega) +
 P^{\dagger}_{{\bf G}{\bf G}^{\prime}}({\bf q},\omega)}{2}
\end{eqnarray}
and
\begin{eqnarray}
B_{{\bf G}{\bf G}^{\prime}}({\bf q},\omega) = 
\frac{P_{{\bf G}{\bf G}^{\prime}}({\bf q},\omega) -
 P^{\dagger}_{{\bf G}{\bf G}^{\prime}}({\bf q},\omega)}{2i},
\end{eqnarray}
then equations (\ref{imaginary_sym}) and (\ref{imaginary_sym2}) still hold,
allowing us to perform  the same computational tasks as for the symmetrized dielectric matrix
 with imaginary frequencies.
This procedure makes it possible  to first compute $W_{{\bf G},{\bf G}^{\prime}}({\bf q},\omega)$
 only for irreducible points of the first Brillouin zone (BZ). We then  determine 
easily  the screened interaction for all {\bf k}-points in the Brillouin zone using symmetry properties.

\subsubsection{Self Energy} 
The self energy $\Sigma$ is 
 the key quantity of any GWA calculation. As previously noticed, we
 have chosen to avoid plasmon-pole models and compute $\Sigma$ with
 the $\omega$ dependence of the screened interaction $W$ within the RPA.\\
\indent First, we split the integral of the selfenergy into a bare exchange or
 Hartree-Fock contribution $\Sigma^{X}$ and an energy dependent
 contribution $\Sigma^{C}(\omega)$ which describes self-energy effects beyond
 $\Sigma^{X}$. The matrix elements of the self-energy are now given 
by the sum of 
\begin{equation}\label{hf_sigma}
\langle \Psi_{{\bf k}n}|\Sigma^{X}|\Psi_{{\bf k}n}\rangle=
-\frac{4\pi}{\Omega}\sum_{\bf q}\sum_{m}^{occ}\sum_{{\bf G}}
\frac{ |M_{{\bf G}}^{mn}({\bf k},{\bf q})|^{2}} {|{\bf q}+{\bf G}|^2}
\end{equation}
where the summation is over occupied states, and \\

$ \langle\Psi_{{\bf k}n}|\Sigma^{C}(\omega)|\Psi_{{\bf k}n}\rangle $
\begin{eqnarray}\label{cor_sigma}
& = &\frac{1}{\Omega}\sum_{\bf{q}}\sum_{{\bf G}{\bf G}^{\prime}}\sum_{m}
\left[M^{mn}_{{\bf G}}({\bf k}, {\bf q})\right]^{\star} 
M^{mn}_{{\bf G}^{\prime}}({\bf k}, {\bf q})
\nonumber \\
& & \times~ C_{{\bf G}{\bf G}^{\prime}}^{m}({\bf k}, {\bf q}, \omega)
\end{eqnarray}
with \\

$ C_{{\bf G}{\bf G}^{\prime}}^{m}({\bf k}, {\bf q}, \omega) $
\begin{eqnarray}\label{C_integrale}
& = &\frac{i}{2 \pi} \int d\omega^{\prime} 
\frac{W^{C}_{{\bf G}{\bf G}^{\prime}}(q,\omega^{\prime})}
{\omega+\omega^{\prime}-\epsilon_m {({\bf k}- {\bf q})}+
i\delta{\rm sgn}(\epsilon_m ({\bf k}- {\bf q})-\mu)}~~~~~~
\end{eqnarray}
where $W^{C}$ is defined as $W^{C}=W-v$, with $v$ being  
the bare Coulomb potential.
To evaluate this  integral directly on the real axis one should  compute $W^{C}$ for
 many points $\omega^{\prime}$ since the shape of $W^{C}$ along the real
 axis is rather ragged. Even though this has been done by some authors,\cite{marini-copper} we
 choose to avoid this difficulty by using the fact that $W^{C}$ is well behaved
 along the imaginary axis. In the present work, we have performed this integral using two 
different methods:\\
\indent In the first one, the contour of the frequency integral (\ref{C_integrale}) is deformed
 in a way to obtain an integral along 
 the imaginary axis plus contributions from the poles of the Green's function.
In this case, we obtain the following expression: \\

$C_{{\bf G}{\bf G}^{\prime}}^{n}({\bf k}, {\bf q}, \omega) $
\begin{eqnarray}\label{C_integrale_bis}
& = & -\frac{1}{\pi} \int_0^\infty d\omega^{\prime \prime} 
W^{C}_{{\bf G}{\bf G}^{\prime }}(q,i\omega^{\prime \prime})
\frac{\omega-\epsilon_n ({\bf k}- {\bf q})}{(\omega-\epsilon_n ({\bf k}- {\bf q}))^2 +\omega^{\prime \prime 2}}. \nonumber \\
& &  \pm W^{C}_{{\bf G}{\bf G}^{\prime}}(q,\pm(\omega-\epsilon_n ({\bf k}- {\bf q})))
\theta(\pm(\omega-\epsilon_n ({\bf k}- {\bf q}))) \nonumber \\
& & ~~~ \times \theta(\pm(\omega-\mu)) \theta(\pm(\epsilon_n ({\bf k}- {\bf q})-\mu)) \nonumber
\end{eqnarray}
The first term represents the contribution along the imaginary axis and is evaluated
 by Gaussian quadrature. The second is from the poles of the Green's function and its computation
 is done by fitting values of $W^{C}$ at $\pm(\omega-\epsilon_n ({\bf k}- {\bf q}))$ from
 values on a given mesh of frequencies\footnote{We have used a simple four  point interpolation method. 
The use of a more elaborate method like the cubic spline  method is shown 
to not influence the final results.}. 
Here  $\mu$ denotes the Fermi level in the LDA and
 $\omega^{\prime \prime}$ is defined to be real.
 This method is similar to the one used by
Aryasetiawan for the implementation of the GWA based on the LMTO method in 
the atomic sphere approximation\cite{review_gw4} (ASA), and within the GWA
of Kotani and coworkers based on  the  full-potential linear muffin-tin orbital  
(FP-LMTO) method.\cite{kotani} 
The reader can find more details about
 this integration  procedure in Refs. \onlinecite{kotani,review_gw4}. 
Similar work has been also carried out by Bechstedt and coworkers\cite{bechstedt} 
 as well as by Fleszar and Hanke\cite{fleszar} starting from a pseudopotential 
approach.  \\
In our second implementation, which is similar to that of Ref.
\onlinecite{Godby},
 we evaluate the matrix elements of the correlative part of the self-energy 
$\langle\Psi_{m{\bf k}}|\Sigma^{C}(\omega)|\Psi_{l{\bf k}}\rangle$ for a set of 
 imaginary frequencies $i\omega$, the resulting quantity is then analytically
 continued to the real axis by fitting it to the following Pad\'e form
\begin{eqnarray}\label{pade_poly}
P(z)= \frac{a_{0}+a_{1}z+a_{2}z^{2}+..+a_{N}z^{N}}{b_{0}+b_{1}z+b_{2}z^{2}+..+b_{M}z^{M}}
\end{eqnarray}
where  $a_{i}$ and $b_{i}$ are complex parameters that  are determined during the fit
 along the imaginary axis. Values of $5$ for N and of $6$ for M provided us
with  an accurate and stable fit. The same kind of continuation has also been
 applied with success to compute  the dynamical response function,\cite{chang-pade1,chang-pade2} 
so we are confident of its reliability.
The main difference between the work presented here and that of Ref. (\onlinecite{Godby}) is that
 our code starts from an all-electron basis, 
so we are not using fast Fourier transforms to switch between
 real and reciprocal spaces and between time and frequency domains.
 Our expression  (\ref{pade_poly}) is also different, but we believe  that this is 
of minor importance\footnote{The reason for not using the same expression as 
in Ref. \onlinecite{Godby} is that our   expression (\ref{pade_poly}) is more general and  
provided us with a better stability 
in the fitting procedure.}.\\
\indent In both cases, the integration over the first Brillouin zone is done by the
special-point technique.\cite{Monkhorst} 
The number of bands as well as the number of $\bf G$ vectors 
in (\ref{cor_sigma}) is increased until the 
QP energies are converged. Similarly, the number of frequency points $\omega^{\prime}$ 
for which $W^{C}$ is computed is increased until 
$C_{{\bf G}{\bf G}^{\prime}}^{n}({\bf k}, {\bf q}, \omega)$  is  well converged.
The two different implementations allow us to check carefully our results, and 
 as can be seen in Table \ref{tab:comparison} for the case of silicon,
the  QP energies are insensitive to the method used to compute the self-energy.


\begin{table}
\caption{\label{tab:comparison}Calculated quasiparticle energies of silicon for some points
 in the Brillouin zone with our two
 different implementations. The results 
are in good agreement with each other. In the last line, the minimum band gap $E_g$ is presented.}
\begin{ruledtabular}
\begin{tabular}{ldd}
 & \multicolumn{1}{c}{First method\footnote{Contour deformation method}}  
& \multicolumn{1}{c}{Second method\footnote{Analytic continuation method}}  \\
\hline
${\bf{\Gamma}}_{1v}$            & -11.85 & -11.87     \\
${\bf{\Gamma}}_{25^{\prime}v}$  & 0.00   & 0.00       \\
${\bf{\Gamma}}_{15c}$           & 3.09   & 3.09       \\
${\bf{\Gamma}}_{2^{\prime}c}$   & 4.05   & 4.06       \\
                                &        &            \\
${\bf X}_{1v}$                  & -7.74  & -7.68      \\
${\bf X}_{4v}$                  & -2.90  & -2.91      \\
${\bf X}_{1c}$                  & 1.01   & 1.03       \\
${\bf X}_{4c}$                  & 10.64  & 10.59      \\
                                &        &            \\
${\bf L}_{2^{\prime}v}$         & -9.57  & -9.50      \\
${\bf L}_{1v}$                  & -6.97  & -6.90      \\
${\bf L}_{3^{\prime}v}$         & -1.16  & -1.17      \\
${\bf L}_{1c}$                  & 2.05   & 2.03       \\
${\bf L}_{3c}$                  & 3.83   & 3.83       \\
                                &        &            \\
$E_g$                           & 0.92   & 0.90       \\
\end{tabular}
\end{ruledtabular}
\end{table}

\subsubsection{Treatment of the Coulomb divergence} 
The last point we wish to discuss is 
an additional difficulty which occurs when evaluating the self-energy by a summation
 of the $\bf q$ points over the full  BZ. We cannot apply the 
special-point technique directly since
 the integrands have a $1/{\bf q}^{2}$ singularity for ${\bf q}
 \rightarrow 0$ as can be seen for example by putting ${\bf G}=0$ in the 
 expression of the exchange term given by (\ref{hf_sigma}). 
The difficulty can be removed by adding and subtracting a term 
which has the same singularities as the initial  expression, and 
which can be evaluated numerically and analytically. As a consequence, 
the  integrals over the BZ are rewritten 
\begin{equation} \label{equa_div}
\sum_{{\bf q}}G({\bf q})=\sum_{{\bf q}} \left[G({\bf q})-A~F({\bf q})\right] 
+ A \sum_{{\bf q}} F({\bf q}), 
\end{equation}
where $F(q)$ is an auxiliary periodic function that diverges like $1/{\bf q}^{2}$ as $\bf q$ vanishes.
 The term is regular and can be evaluated by the special point technique 
whereas the last sum  is evaluated analytically.
For the exchange term, it is not difficult to evaluate $A$ in (\ref{equa_div}), 
 but it gets more complicated for the correlative part of the self-energy 
(\ref{cor_sigma}). The purpose of the offsetted $\Gamma$-point method\cite{kotani} 
is to avoid the evaluation of the quantity $A$, but still to be able to deal with
 the divergence. The main idea is to find a new mesh of points such that 
\begin{equation} \label{fq_egal}
\sum_{{\bf q}} F({\bf q}) = \sum_{{\bf q^\prime}} F({\bf q^\prime})
\end{equation}
where the $\Gamma$-point is included in the old mesh ${\bf q}$ but not 
 in the new one ${\bf q^\prime}$: the $\Gamma$-point is replaced by other points
(different from $\Gamma$) to construct the $\bf q^\prime$ grid in order to 
 fulfill Eq. (\ref{fq_egal}). Eq. (\ref{equa_div}) is therefore rewritten as:
\begin{equation} \label{equa_div2}
\sum_{{\bf q}}G({\bf q})=\sum_{{\bf q}} \left[G({\bf q})-A~F({\bf q})\right] 
+ A \sum_{{\bf q^\prime}} F({\bf q^\prime}), 
\end{equation}
Then we show by inspection  that the term 
 $\sum_{{\bf q}} \left[G({\bf q})-A~F({\bf q})\right] $ is equal to 
$\sum_{{\bf q^\prime}} \left[G({\bf q^\prime})-A~F({\bf q^\prime})\right] $ with
a controlled error, Eq. (\ref{equa_div2}) transforms to
\begin{eqnarray} \label{equa_div_new}
\sum_{{\bf q}}G({\bf q}) &= & \sum_{{\bf q^\prime}} G({\bf q^\prime})
\end{eqnarray}
because the two terms which contain the function $F({\bf q})$ cancel out since
they are evaluated on the same ${\bf q^\prime}$ grid. We have therefore avoided
 the evaluation of the complicated $A$ term in Eq. (\ref{equa_div}).

The remaining points to be addressed are the choice of the function $F({\bf q})$ 
and the number of additional points for the new mesh used to solve (\ref{fq_egal}).
In our case, we write $$F({\bf q}) = \sum_{{\bf G}}\frac{\text{exp}(-\mid \bf{q}+\bf{G}\mid^{2})}
{\mid \bf{q}+\bf{G}\mid^{2}}$$ and choose to add $6$ points to the original ${\bf{q}}$ mesh 
in order to get the new mesh. Equation (\ref{fq_egal}) is then solved to provide
us with the coordinates of the new $6$ points in the BZ.
 \footnote{To justify the
 use of the present expression for $F({\bf q})$, we have also used
the function given by Ref. \onlinecite{Gygi} and found that the final QP 
energies differ by
 at most  0.02 eV for silicon. Moreover, the $F({\bf q})$ is specific to
 fcc lattice systems and therefore must be adapted for other systems according
 to Ref. \onlinecite{Gygi-general}, whereas the function used in this work is independent
 of crystallographic system.
 We have also checked if $6$ points are sufficient by performing calculations
 using  a set of $12$ points, the resulting QP energies remain unchanged, proving  the
 validity of our choice.}
 The computational cost is further
 reduced by finding equivalent points among those $6$ points, and we  end up 
 with well behaved and easily evaluated BZ sums.

\section{\label{sec:three} Numerical results and discussion}
In this section we present our theoretical quasiparticle energies for
 the six materials studied in this paper, together with the available
theoretical and experimental results. In Sec. \ref{subsec:three-one} we report our results
 for four semiconductors (Si, SiC, AlAs, InAs) of zinc-blende type structure, 
 while Sec. \ref{subsec:three-two} is devoted to studying the alkali hydrides NaH and KH in 
the rocksalt phase.  Table \ref{tab:cell_parameter} presents the experimental lattice 
parameters and the energy cut offs $E_{cut}$ used for the final converged calculations. 
\begin{table}
\caption{\label{tab:cell_parameter}
Lattice constants $a$ (in atomic units) and energy cut offs $E_{cut}$  (in Rydberg) used
 for our PAW calculations. The lattice constants are from Ref. \onlinecite{book_lat}, unless
 stated otherwise.}
\begin{ruledtabular}
\begin{tabular}{ldc}
                          &\multicolumn{1}{c}{$a$}   & \multicolumn{1}{c}{E$_{cut}$} \\ \hline
Si                        & 10.26 & 20    \\
SiC                       & 8.24  & 25    \\
AlAs                      & 10.67 & 20    \\
InAs                      & 11.41 & 20    \\
NaH                       & 9.28\footnotemark[1]  & 40   \\
KH                        & 10.83\footnotemark[1]  & 40   \\
\end{tabular}
\end{ruledtabular}
\footnotetext[1]{Reference \onlinecite{martins}}
\end{table}

\subsection{\label{subsec:three-one}Results for Si, SiC, AlAs, and InAs} 
\begin{figure}[h]
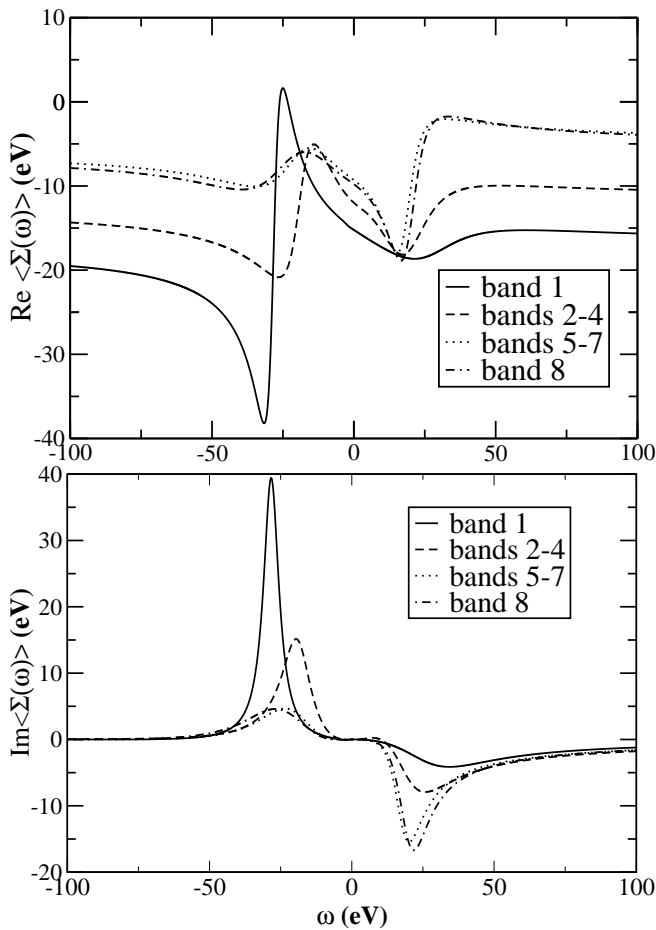

\includegraphics*[angle=0,width=0.48\textwidth]{fig1.eps}
\includegraphics*[angle=0,width=0.48\textwidth]{fig2.eps}
\caption{
\label{fig:self-godby} 
Re$\langle\Psi_{{\bf k}m}|\Sigma(\omega)|\Psi_{{\bf k}m}\rangle $ and 
Im$\langle\Psi_{{\bf k}m}|\Sigma(\omega)|\Psi_{{\bf k}m}\rangle $
shown for the first 8 bands for silicon at $\bf{k}=0$. 
The zero of energy is at the center of the band gap.}
\end{figure}

Silicon is probably the most carefully studied semiconductor, and several GWA results are available. 
Using silicon as a prototype will  allow us 
to test our method by making careful comparisons with previous GWA calculations. 
As mentioned earlier, our code presents
 two different ways for calculating the self-energy. We therefore test their accuracy  
for silicon  in Table \ref{tab:comparison}.  We find that 
the results of the two methods are almost identical,
 showing that they  are equally reliable for computing 
 the self-energy. In particular, our implementation with the 
 extrapolation procedure makes it possible  to represent the full-frequency dependence of the
 self-energy with a small additional computational cost. 
Fig. \ref{fig:self-godby} shows 
 the real and imaginary  parts of $\Sigma$ along the real axis for silicon at the 
$\Gamma$ point for a wide range of frequencies.
 The agreement with previous work is excellent.\cite{Godby} 

A special feature of our
 work is the possibility of obtaining the imaginary part of the
 self-energy (see Fig. \ref{fig:self-godby}), a task virtually impossible 
within the PlP approximation\footnote{The imaginary
 part of the self-energy is a sum of delta functions in the plasmon-pole approximation.}. 
The spectral function which can be obtained directly from the self-energy\\

\begin{figure}[h]
\includegraphics*[angle=0,width=0.40\textwidth]{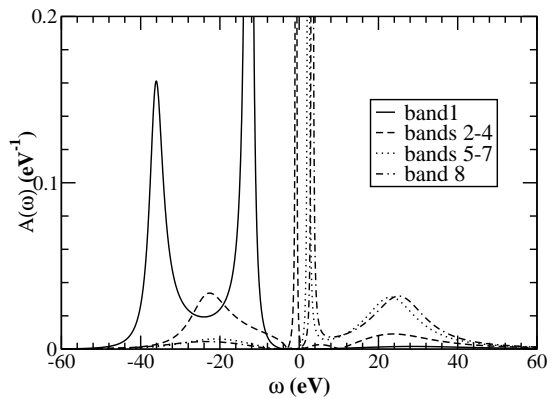}
\caption{\label{fig:spectral-godby} 
Spectral function $\langle\Psi_{{\bf k}m}|A(\omega)|\Psi_{{\bf k}m}\rangle$ 
for the first 8 bands for silicon at $\bf{k}=0$.
The zero of energy is at the center of the band gap.
The sharp peaks are QP poles, their weights correspond to the factor Z
 defined in Eq. \ref{Renormalization}.
}
\end{figure}

$\langle\Psi_{{\bf k}m}|A(\omega)|\Psi_{{\bf k}m}\rangle =$
\begin{eqnarray}\label{spectral_function}
\frac{|\text{Im}\langle\Psi_{{\bf k}m}|\Sigma(\omega)|\Psi_{{\bf k}m}\rangle|}
{[\omega-\epsilon_m({\bf k})-\text{Re}\langle\Psi_{{\bf k}m}|\Sigma(\omega)|\Psi_{{\bf k}m}\rangle]^{2} 
+ [\text{Im}\langle\Psi_{{\bf k}m}|\Sigma(\omega)|\Psi_{{\bf k}m}\rangle]^{2}} \nonumber
\end{eqnarray}
is of major interest since it can be used for the interpretation of  
 experimental  photoemission and inverse-photoemission spectra. As an example, the 
spectral function of silicon at the $\Gamma$ point is shown in Fig. \ref{fig:spectral-godby}.
The sharp peaks correspond to QP excitations, while the incoherent 
part of the function, the spectral background, is much complicated and could correspond
to plasmon type excitations. 

\begin{table*}
\caption{\label{tab:silicon}Selected energy eigenvalues, in eV, at $\Gamma$, $X$ and $L$ for
 Si. Our results are compared with two other all-electron implementations 
of the GW method and
 with experimental results. Data in parentheses are results when the denominator of the Green's function
 is updated with QP energies. Data in the last line correspond to the minimum energy gap $E_g$.}
\begin{ruledtabular}
\begin{tabular}{ldddddcc}
 & \multicolumn{2}{c}{LDA} & \multicolumn{3}{c}{GW approximation}     & Expt.\footnotemark[3] \\
 & \multicolumn{1}{c}{Present} & \multicolumn{1}{c}{LAPW\footnotemark[1]} & \multicolumn{1}{c}{~~~~~~~~Present} & \multicolumn{1}{c}{PAW-PlP\footnotemark[2]} &  \multicolumn{1}{c}{LAPW\footnotemark[1]}  & \\ \hline

${\bf{\Gamma}}_{1v}$            & -11.97  & -11.95  & -11.85 ~(-11.89) & -11.92 & -12.21  & -12.5$\pm$0.6\\
${\bf{\Gamma}}^{\prime}_{25v}$  & 0.00  & 0.00    & 0.00 ~(0.00) & 0.00 & 0.00&      0.00 \\
${\bf{\Gamma}}_{15c}$           & 2.54  & 2.55    & 3.09 ~(3.15) & 3.16 & 3.30&   3.40,3.05\footnotemark[4] \\
${\bf{\Gamma}}^{\prime}_{2c}$   & 3.23  & 3.17    & 4.05 ~(4.12) & 4.09 & 4.19&   4.23, 4.1\footnotemark[4] \\
                                &           &              &                    \\
${\bf X}_{1v}$                  & -7.82 & -7.82    & -7.74 ~(-7.78) & -7.91 &-8.11  &                     \\
${\bf X}_{4v}$  & -2.85 & -2.84 & -2.90 ~(-2.92) & -2.98 &-3.03  & -2.9\footnotemark[5], -3.3$\pm$0.2\footnotemark[6]\\
${\bf X}_{1c}$                  &  0.61 & 0.65    & 1.01 ~(1.08) & 1.10 &1.14&   1.25\footnotemark[4]  \\
${\bf X}_{4c}$                  & 10.02  &    & 10.64  ~(10.72) & 10.74 &  &               \\
                                &           &               &                \\
${\bf L}^{\prime}_{2v}$         & -9.63 & -9.63    & -9.57 ~(-9.60)& -9.66 &-9.92 &   -9.3$\pm$0.4\\
${\bf L}_{1v}$                  & -6.99 & -6.98    & -6.97 ~(-7.00)& -7.15 &-7.31  &  -6.7$\pm$0.2  \\
${\bf L}^{\prime}_{3v}$ & -1.19 & -1.19    & -1.16 ~(-1.17)& -1.24 &-1.26 & -1.2$\pm$0.2  \\
${\bf L}_{1c}$      &  1.44 & 1.43    & 2.05 ~(2.11) & 2.08 &2.15 & 2.1\footnotemark[7],2.4$\pm$0.15  \\
${\bf L}_{3c}$       &  3.30 & 3.35    & 3.83 ~(3.90) & 3.92 &4.08  & 4.15$\pm$0.1\footnotemark[8] \\
                                &           &              &                    \\
$E_g$                           &  0.55 & 0.52 & 0.92 ~(0.95) & 0.97 & 1.01 & 1.17 \\
\end{tabular}
\end{ruledtabular}
\footnotetext[1]{Reference \onlinecite{Hamada}}
\footnotetext[2]{Reference \onlinecite{brice}}
\footnotetext[3]{Unless noted, Ref \onlinecite{book_lat}}
\footnotetext[4]{Reference \onlinecite{Ortega}}
\footnotetext[5]{Reference \onlinecite{Spicer}}
\footnotetext[6]{Reference \onlinecite{Wachs}}
\footnotetext[7]{Reference \onlinecite{Hulthen}}
\footnotetext[8]{Reference \onlinecite{Himpsel1}}

\end{table*}
\begin{table*}[h]
\caption{\label{tab:others} Quasiparticle energies in eV  at $\Gamma$, $X$ and $L$ for
SiC, AlAs, and  InAs. 
Data in parentheses are results when the denominator of the Green's function
 is updated with QP energies.
Our results are compared with PP-GW method 
and with experimental results (minimum band gaps are underlined).}
\begin{ruledtabular}
\begin{tabular}{lccccc}
   & \multicolumn{2}{c}{LDA} & \multicolumn{2}{c}{GW} & \multicolumn{1}{c}{Expt\footnotemark[1]} \\
   & Present & PP\footnotemark[2] & Present & PP\footnotemark[2] &\\ \hline
SiC &  & & & &\\
${\bf{\Gamma}}^{\prime}_{25v}\rightarrow {\bf{\Gamma}}_{15c} $ & 6.25 & 6.41 & 7.32(7.45) & 7.35 & \\
${\bf{\Gamma}}^{\prime}_{25v}\rightarrow {\bf X}_{1c} $ & \underline{1.28}& \underline{1.31} & \underline{1.80(1.89)} & \underline{2.34} & \underline{2.39}\\
${\bf{\Gamma}}^{\prime}_{25v}\rightarrow {\bf L}_{1c} $     & 5.34 & 5.46 & 6.45(6.56) & 6.53 & 4.2 \\
AlAs &  & & & &\\
${\bf{\Gamma}}^{\prime}_{25v}\rightarrow {\bf{\Gamma}}_{15c} $ & 1.94 & 1.77 & 2.72(2.79) & 2.75 & 3.11\footnotemark[3]\\
${\bf{\Gamma}}^{\prime}_{25v}\rightarrow {\bf X}_{1c} $  &\underline{1.32} & \underline{1.20} & \underline{1.57(1.65)} & \underline{2.08} & \underline{2.24}\\
${\bf{\Gamma}}^{\prime}_{25v}\rightarrow {\bf L}_{1c} $ & 2.06 & 1.89 & 2.73(2.80) & 2.79 & 2.49\footnotemark[3];2.54\footnotemark[4]\\
InAs &  & & & &\\
${\bf{\Gamma}}^{\prime}_{25v}\rightarrow {\bf{\Gamma}}_{15c} $ & \underline{-0.07}  & \underline{-0.39} & 
\underline{0.46(0.49)} & \underline{0.59} & \underline{0.60}\\
${\bf{\Gamma}}^{\prime}_{25v}\rightarrow {\bf X}_{1c} $        & 1.48 &       & 1.57 (1.61) & 2.10 &      \\
${\bf{\Gamma}}^{\prime}_{25v}\rightarrow {\bf L}_{1c} $        & 1.05 &       & 1.54 (1.58) & 1.52 & 1.74\\
\end{tabular}
\end{ruledtabular}
\footnotetext[1]{Unless noted Ref \onlinecite{book_lat}. For InAs, data have been averaged to account
 for the neglect of spin-orbit coupling in our case.}
\footnotetext[2]{For SiC, PP results are from Ref. \onlinecite{Rohlfing2}, for AlAs from Ref.
 \onlinecite{shirleylouie2}, for InAs from Ref. \onlinecite{zhulouie} and averaged to account
 for the neglect of spin-orbit splitting.}
\footnotetext[3]{Reference \onlinecite{Wolford}.}
\footnotetext[4]{Reference \onlinecite{Aspnes}.}
\end{table*}
The QP calculations have been performed using 256 $\bf{k}$ points in the full BZ. The
 size of the dielectric matrix defined in Eq. (\ref{symetrised_epsilon}) is $137 \times 137$ 
for silicon and SiC, $169 \times 169$ for AlAs, $181 \times 181$ for InAs. 
$200$ bands were used for the sum over conduction states in Eq. (\ref{symetrised_epsilon})
 and for the sum over $m$ in Eq. (\ref{cor_sigma}). Due to the smoothness of the integrand 
along the imaginary axis, 11 points are found sufficient to obtain well converged quantities.
An energy  step of $1.5$ eV is used for the part of Eq. (\ref{C_integrale_bis}) which corresponds 
to the poles of the Green's function. Using this energy step we determine an energy grid which 
we use to produce an accurate fit to 
$W^{C}_{{\bf G}{\bf G}^{\prime}}(q,\pm(\omega-\epsilon_n ({\bf k}- {\bf q})))$ for 
the different points $\pm(\omega-\epsilon_n ({\bf k}- {\bf q}))$. 
All these high values of the parameters ensure the convergence of the  QP energies to within  $0.05$ eV.

Table \ref{tab:silicon} shows the excellent agreement of our results with two other all-electron 
GWA implementations of the QP energies of silicon.
From this table it seems that, at least for Si, the overall difference between the
RPA and the PlP results is small.\footnote{Notice that the same PAW method is used to 
compute the Si QP energies both within the RPA and the PlP model.}. 
Nevertheless,  a discrepancy of as much as  $0.18$ eV
 for the energy of ${\bf L}_{1v}$ is obtained. It seems then, at least for Si, the PlP model 
overestimates only slightly the differences between the energy levels within the
GWA. 

Table \ref{tab:others} compares the calculated QP energies for 3C-SiC (also known
 as $\beta$-SiC), AlAs and InAs with experimental data as well as with pseudopotential-GWA (PP-GWA) 
calculations. 
 The band gaps are given at $\Gamma$, $X$, and $L$ and are underlined in this table. 
These studies are motivated by the
fact that SiC is a material of high current technological interest and that 
InAs is predicted to be metallic in the LDA,  whereas AlAs is used as a test 
case. 
A general trend of our implementation is that the agreement with experiment as well as with PP-GWA 
 results is not perfect, as also found by other implementations based on all-electron 
methods;\cite{kotani,brice} reporting differences up to $0.4$ eV. 
In particular, the largest difference occurs for the 
${\bf{\Gamma}}^{\prime}_{25v}\rightarrow {\bf X}_{1c} $ 
transition.\footnote{Inspection of results of Table \ref{tab:others} shows that the 
 ${\bf{\Gamma}}^{\prime}_{25v}\rightarrow {\bf L}_{1c} $ for SiC seems to be largely overestimated 
by GW calculations.  In fact, we join the conclusion of Ref. \onlinecite{Rohlfing2} and claim that 
 the experimental value is certainly less precise. This is confirmed by the fact that our calculation 
 agrees by about $0.08$ eV with their data.} 
We showed in a previous study\cite{brice} for the case of silicon that 
these differences are mainly traced back to differences
 between the exchange-correlation matrix elements obtained by the two methods. 
We believe that  this can be extended to other materials as well, since 
it seems to be a general feature
\footnote{Using the second method presented in this paper (i. e., the analytic continuation method) did
 not change the value of the ${\bf{\Gamma}}^{\prime}_{25v}\rightarrow {\bf X}_{1c} $ transition by more
 than $0.05$ eV.} of all-electron GWA calculations. \\ 
\indent In fact, the difference between 
 all-electron and PP based GW calculations is not surprising since the use of pseudo-wave functions
 for evaluating matrix elements of a  general operator produce results that may not be sufficiently precise, 
because the  pseudo-wave functions
 are constructed to reproduce the all-electron wave functions only in the 
interstitial region, i.e.,  outside the atomic 
spheres. This is however a good representation for studying  
properties that depend only on the behavior of the wave function in the bonding region.
An error is then introduced in any PP-GWA calculation.  This error  seems to fortuitously 
have a tendency to improve the  agreement with experiment,  
explaining the exceptional success of the PP-GWA.\\
\indent For InAs, the incorrect metallic behavior obtained within the LDA 
is corrected by our GWA calculation. The GWA 
produces the true semiconducting 
 state as given by experiment. Since we did not account for  the spin-orbit coupling  in our calculation,
 we have simply averaged  out the spin-orbit-split experimental  values to make the comparison
 with our work possible. \\

\subsection{\label{subsec:three-two}Results for the alkali hydrides NaH and KH} 

Alkali hydride materials exhibit a structural phase transition
from the B1 (NaCl structure)  to the B2 (CsCl structure) type structure under hydrostatic pressure. 
A number of studies have been performed to understand the equation of state of these 
materials,\cite{rodriguez,kunz,martins,ahuja}
as well as the possibility of an insulator-metal transition,\cite{hama,kulikov} however 
only  few studies  have been published about the electronic structures.\cite{rodriguez,kunz,database}
In these materials, hydrogen behaves as an $H^{-}$ ion, leading to partly ionic materials 
with a larger band gap than the  studied '$sp$' semiconductors.  It is, therefore, of interest 
to know whether the GWA is capable of producing such large band gaps. 
In this study we are only concerned with the determination of their QP energies
for the rocksalt crystallographic structure.  \\ 
\begin{figure}[h]
\includegraphics[width=0.45\textwidth]{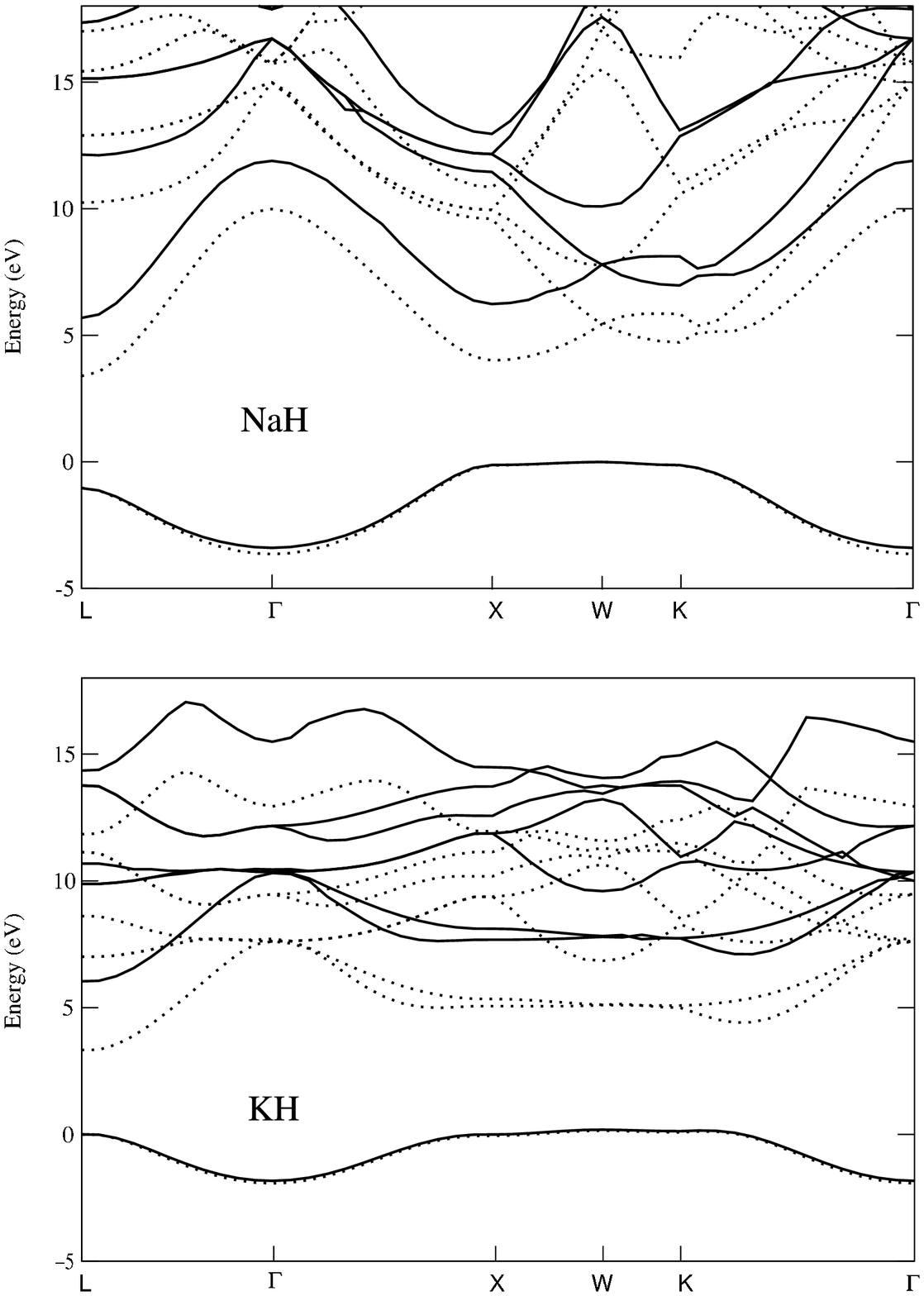}
\caption{\label{fig:band2} Calculated LDA (doted lines)  and GW (full lines)  
electronic band structures of  NaH and KH along some high-symmetry directions.
These calculations are performed   using  the parameters reported in 
Table \ref{tab:cell_parameter}.}
\end{figure}
\begin{table}[h]
\caption{\label{tab:nah} Quasiparticle energies in eV  at $\Gamma$, $X$, $L$, $W$, and $K$ for
 NaH and KH. The last line shows the minimum band gap $E_g$.
Data in parentheses are results when the denominator of the Green's function
 is updated with QP energies. We are not aware of any experimental study concerning the 
electronic band structure of these  materials.}
\begin{ruledtabular}
\begin{tabular}{lccccc}
   & \multicolumn{2}{c}{NaH} && \multicolumn{2}{c}{KH} \\
                          & \multicolumn{1}{c}{LDA} & \multicolumn{1}{c}{GW}
                          && \multicolumn{1}{c}{LDA} & \multicolumn{1}{c}{GW} \\
\hline
${\bf{\Gamma}}_{1v}$            & -3.64&  -3.39 ~(-3.59) && -2.07&   -2.02 ~(-2.11)     \\
${\bf{\Gamma}}_{1c}$            & 9.98&   11.89 ~(12.38) && 7.47&   9.81 ~(10.43)      \\
${\bf{\Gamma}}_{25^{\prime}c}$  & 14.98&  16.71 ~(17.24) && 7.57&   10.16  ~(10.74)     \\
${\bf{\Gamma}}_{15c}$           & 15.72&  17.87 ~(18.11) && 9.33&   11.98 ~(12.53)     \\
                                &          &   &&    &         \\
${\bf X}_{1v}$                  & -0.13&  -0.12 ~(-0.12) && -0.18& -0.18 ~(-0.20)      \\
${\bf X}_{4^{\prime}c}$         & 4.00&   6.23 ~(6.66)   && 4.91&  7.49 ~(8.01)      \\
${\bf X}_{3c}$                  & 9.59&   11.45 ~(11.96) && 5.18&  7.92 ~(8.47)      \\
${\bf X}_{5^{\prime}c}$         & 9.95&   12.15 ~(12.60) && 8.76&  11.68 ~(12.24)      \\
${\bf X}_{1c}$                  & 10.84&  12.95 ~(13.36) && 9.27&  12.31 ~(12.86)       \\
                                &          &   &&    &         \\
${\bf L}_{1v}$                  & -1.03 & -1.03 ~(-1.06) && -0.15 & -0.19 ~(-0.20)      \\
${\bf L}_{2^{\prime}c}$         & 3.39  & 5.68 ~(6.11)   && 3.18 & 5.85 ~(6.35)       \\
${\bf L}_{1c}$                  & 10.24 & 12.13 ~(12.59) && 6.86 & 9.69 ~(10.23)       \\
${\bf L}_{3^{\prime}c}$         & 12.89 & 15.14 ~(15.61) && 8.46  & 10.49 ~(11.18)       \\
${\bf L}_{3c}$                  & 15.42 & 17.34 ~(17.77) && 10.98 & 13.57 ~(14.03)        \\
                                &          &   &&    &         \\
${\bf W}_{1v}$                 & 0.00 &   0.00 ~(0.00)   && 0.00 & 0.00 ~(0.00)        \\
${\bf W}_{3c}$                  & 5.43&   7.79 ~(8.24)   && 4.96 & 7.60 ~(8.13)      \\
${\bf W}_{2^{\prime}c}$         & 7.75&   10.08 ~(10.55) && 6.70 & 9.40 ~(9.94)     \\
${\bf W}_{1c}$                  & 15.50&  17.57 ~(17.98) && 10.50 & 13.03 ~(13.55)    \\
${\bf W}_{3c}$                  & 17.08&  19.16 ~(19.57) && 10.73 & 13.25 ~(13.78)     \\
                                &          &   &&    &         \\
${\bf K}_{1v}$                  & -0.13   & -0.12 ~(-0.12)        && -0.06  & -0.06 ~(-0.07)     \\
${\bf K}_{3c}$                  & 4.71 &    6.96 ~(7.41)          && 4.84    &  7.50  ~(8.02)     \\
${\bf K}_{1c}$                  & 5.84 &    8.11 ~(8.56)          && 4.94    &  7.62  ~(8.18)      \\
${\bf K}_{4c}$                  & 10.57 &   12.85 ~(13.27)        && 8.13  &  10.72 ~(11.38)    \\
${\bf K}_{1c}$                  & 11.02 &   13.09 ~(13.52)        && 8.36  &  10.94 ~(11.54)     \\
                                &          &   &&    &         \\
$E_g$    &          3.39  & 5.68 ~(6.11)  &&   3.18     & 5.85~(6.35)     \\
\end{tabular}
\end{ruledtabular}
\end{table}
In Fig. \ref{fig:band2} we report the QP band structures of NaH and KH  
along some high-symmetry directions,\footnote{For the calculated electronic properties of NaH 
and KH alkali hydrides we have used $64$ {\bf k}-points in the full BZ  as well as $200$ bands, and a 
dielectric matrix of size  $169 \times 169$ to achieve well converged results.} 
and present a detailed overview of their LDA and QP energies in Table \ref{tab:nah}. 
In the following we detail and compare our results with existing experimental and calculated
results:  
For NaH, the authors of Ref. \onlinecite{rodriguez} reported  an LDA calculation with  
 an improved   LMTO method in its atomic sphere approximation, however, 
from their band structure plot we estimated that their band gap is only about $2.7$ eV and
is  direct at the ${L}$ point. This disagrees with our calculation, since 
at the LDA level we found an indirect band gap of $3.39$ eV from 
${ W}$ to ${ L}$. This could be   due to Ref. \onlinecite{rodriguez}'s use 
of a different value of the lattice 
parameter of $8.90$ a.u.  A calculation by Kunz and Michish\cite{kunz} 
based on the electronic polaron model produced a direct band gap at ${ X}$ of $1.52$ eV,
a value too low for these ionic materials, so  that it should  be taken only at a qualitative level.
\footnote{Kunz and Mickish
have also reported a band gap for LiH of 6.61 eV compared to a measured value of 4.99 eV
(data reported in the paper of S. Baroni {\sl et al.}, Phys. Rev. B {\bf 32}, 4077 (1985)).} 
However, our LDA results are  in  full agreement with results of 
Ref. \onlinecite{database}, where  an indirect band gap from $W$ 
to $L$ of about $3.3$ eV was found.
Table \ref{tab:nah} shows that the energy level differences of NaH are substantially 
increased by the use of the GWA compared with the LDA results. 
In particular, the minimum band gap is $5.68$ eV within the GWA, whereas
 it's only $3.39$ eV within the LDA. 
An other interesting point is that  we found that the most used  scissors-operator shift
 seems  not to apply for the computation of optical properties of NaH. 
This is because the band gap is increased by 1.66 eV at  the $\Gamma$ point, and as much as 
 2.34 eV for the direct transition at ${L}$ point. 
A GWA calculation of the QP energies in the full BZ  is then required for 
the  study the  optical properties of NaH. 
Table \ref{tab:nah} shows also our  LDA and QP results for potassium 
hydride.  Our LDA results are in good agreement with the calculation in 
Ref. \onlinecite{database}.
We find that in both LDA and GWA calculations, the band gap is indirect from $W$ to $L$, and
 could easily switch to a direct gap with a small lattice parameter variation since the
 valence band (1s state of hydrogen) is very flat, i.e., the  valences states at the ${X}$, 
${L}$, ${W}$, and  ${K}$ have the same energies within  $0.2$ eV accuracy. 
The previous remark about the non-validity of the scissors-operator shift 
 still holds here for the same reasons.  
The LDA band gap increases by an amount of $2.29$ eV at $\Gamma$, and as much as  
$2.71$ eV at the ${\bf L}$ point.  \\
\indent It is surprising   
to notice that, across the whole Brillouin zone, the GW band gap shift of KH is  larger than that 
of NaH  despite that the LDA band gap of NaH is larger than that of KH.  The reason for this 
puzzling large shift is that the screening of the Coulomb potential in KH is found to be  less 
efficient than in NaH. Indeed, we have found that the RPA static dielectric function of NaH is 
3.43  much larger than that of KH which is about 2.62. The hybridization is also less strong
for KH than for NaH, since the band width of hydrogen $s$-states is about 2.02 eV  for KH (much 
smaller than the  3.64 eV band width of NaH).
As a consequence,  the higher excited states are lower in energy, across the Brillouin zone, 
by about 2 to 5 eV for  KH than for NaH.
 We hope that our predictive results will stimulate 
experimentalists to perform photoemission or  optical studies
of  these interesting materials.
\section{\label{sec:four} Conclusion}
We have presented an implementation of the GWA using the all-electron PAW 
method where the screened Coulomb interaction is obtained using the RPA  
dynamical dielectric function. Thus we avoided  the use of the 
 plasmon-pole approximation. 
We have applied it to study the QP energies of
 Si, SiC, AlAs, and InAs and found  that a precise 
comparison with other available theoretical and
experimental results shows that sometimes the GWA can lead to noticeable discrepancies  
 with experiments. 
Those discrepancies  are generally not so pronounced in the pseudopotential GWA 
calculations using PlP models. We argued that the approximations used in the 
pseudopotential method have a tendency to fortuitously  improve  the agreement with 
experiment. \\
\indent We have presented  detailed results  for the first time for NaH and KH alkali hydrides,
 and showed that the GWA enhanced substantially the LDA band gaps, 
motivating further theoretical and experimental  studies. \\
\indent Since our method can compute the imaginary part of the self-energy, 
we could then determine   the QP lifetimes, a task not possible 
using the  PlP approximation. 
Further inspection of spectral properties as well
as the computation of QP lifetimes will be presented in
future work. The method is currently being  applied to determine the
excitation properties of LiH, and  the results  will be reported 
elsewhere.\cite{seb_lifetime}
Moreover, the use of symmetry and an efficient
 implementation make us confident that we will soon be able to 
study  systems with  a large  number of atoms per unit cell, like surfaces 
or polymers.
Furthermore, because  we use a mixed   basis-set in our implementation we could 
 study systems with localized '$d$' or '$f$' electrons with a reduced computational 
cost compared with methods based  only a plane wave basis-set. \\
\begin{acknowledgments}
One of us (S. L.) is particularly grateful to W. E. Pickett since part of this
 work was done during a visit to the University of California Davis
supported by DOE grant DE-FG03-01ER45876. 
Supercomputer time was provided by CINES (project gem1100) on the IBM SP3.
This research was supported in part by the National Science
Foundation under Grant No. PHY99-07949.
\end{acknowledgments}


\end{document}